\documentclass[twocolumn,epsfig,superscriptaddress,prl]{revtex4-2}

\usepackage{graphicx}
\usepackage{amssymb}
\usepackage{amsmath}
\usepackage{hyperref}
\usepackage{changes}
\usepackage{comment}

\begin{document}
\title{Microscopic theory of the current-voltage characteristics of Josephson tunnel junctions}	

\author{Sang-Jun Choi}
\email{sang-jun.choi@physik.uni-wuerzburg.de}
\affiliation{Institute for Theoretical Physics and Astrophysics, University of W\"urzburg, D-97074 W\"urzburg, Germany}
\author{Bj\"orn Trauzettel}
\affiliation{Institute for Theoretical Physics and Astrophysics, University of W\"urzburg, D-97074 W\"urzburg, Germany}
\affiliation{W\"urzburg-Dresden Cluster of Excellence ct.qmat, Germany}
\date{\today}

\begin{abstract}
{Deep theoretical understanding of the electrical response of Josephson junctions is indispensable regarding both recent discoveries of new kinds of superconductivity and technological advances such as superconducting quantum computers. Here, we study the microscopic theory of the DC current-biased $I$-$V$ characteristics of Josephson tunnel junctions. We derive an analytical formula of the $I$-$V$ characteristics of generic junctions. We identify subharmonics of the $I$-$V$ characteristics and their underlying mechanism as the feedback effect of intrinsic AC currents generated by voltage pulses in the past. We apply our theory to analytically solve the Werthamer equation and describe various DC current-biased $I$-$V$ characteristics as a function of softening of the superconducting gap. Strikingly, we identify voltage staircases of the $I$-$V$ characteristics in a genuine Josephson junction without AC current bias or qubit dynamics. Our general analytical formalism opens new avenues for a microscopic understanding of $I$-$V$ characteristics of Josephson junctions that have been limited to phenomenological models so far. }
\end{abstract}
\maketitle

Microscopic theories of Josephson tunnel junctions were established at the early stages of the discovery of the Josephson effect~\cite{Cohen,Josephson1962,Ambegakokar1963,Werthamer,Larkin}. 
For the DC voltage-biased junctions yielding intriguing phenomena such as multiple Andreev reflections~\cite{Josephson1962,Klapwijk,Bratus,Averin,Zaitsev,Naveh}, the $I$-$V$ characteristics can be well understood with the simple equation of motion of the superconducting phase difference $\phi(t)$ with $\dot{\phi}(t)=2eV/\hbar$ at voltage bias $V$. However, DC current-biased junctions
are governed by a nonlinear and  nonlocal-in-time integro-differential equation, and a microscopic theory of the DC current-biased $I$-$V$ characteristics of Josephson junctions requires to solve the complex dynamics of $\phi(t)$. 

While the $I$-$V$ characteristics of DC current-biased Josephson junctions are widely examined experimentally, its theoretical analysis has been limited to phenomenological theories~\cite{McCumber,Stewart,Scott,Schlup}, adiabatic approximations~\cite{Larkin,Schlup}, or numerical approaches~\cite{McDonald, Schlup1978, Zorin}, in which several drawbacks are faced. The analytical theories on the $I$-$V$ characteristics are typically based on reducing the nonlocal-in-time integro-differential governing equation of $\phi(t)$ into a local-in-time differential equation. However, those theories, especially at low temperature, are self-consistent only for zero voltage $V=0$ or in the Ohmic regime $V\gg I_cR_n$. The inconsistency of the analytical theories at intermediate voltages, $0<V<I_cR_n$, is related to the fast dynamics of $\phi(t)$ resulting in voltage pulses~\cite{Schlup1977}. These voltage pulses are inevitable due to the nonlinear nature of the supercurrent~\cite{Strogatz}. While numerical approaches have been successfully implemented in the frequency-domain, they have mainly been performed for conventional BCS superconductors. Thus, it is imperative to put forward a rigorous analytical theory, which is consistent and generally applicable to various types of current-biased Josephson junctions.
 
In this Letter, we study the microscopic theory of DC current-biased  $I$-$V$ characteristics of generic Josephson tunnel junctions. 
The gist of our approach is to analyze the general behavior of the superconducting phase $\phi(t)$ generated by successive voltage pulses in the time domain. With this approach, we show that the voltage pulses dynamically modulate the tilted washboard potential in the RSJ model. This viewpoint provides us with a clear understanding of the complex dynamics of $\phi(t)$. Focusing on an intermediate voltage strength across the junction, we find that the retarded response of the Josephson junctions is decisive for the $I$-$V$ characteristics. This enables us to obtain an analytical formula of the $I$-$V$ characteristics for general memory kernels in arbitrary Josephson tunnel junctions.

We apply our theory to the seminal example of the Werthamer equation including smearing of the quasiparticle density of states (DOS). We emphasize the validity of our approach by a convincing agreement between numerics and our analytical formula. If the smearing is small, we predict voltage staircases in DC current-biased $I$-$V$ characteristics resembling Shapiro steps that occur if an additional AC current bias is applied to the junction. Finally, we show that a substantial smearing of the DOS can justify the validity of the RSJ model at zero temperature but with hystereses due to nonequilibrium Josephson effects. Both voltage staircases and hystereses are experimentally observed in high-quality Josephson junctions~\cite{Gul,Charpentier,Mayer,Ridderbos,Perla}. In the presence of a qubit formed in the junction, voltage staircases and hystereses have recently been predicted in the DC current-biased case as well~\cite{Feng2018,Choi2020,Oriekhov2021}. In a genuine Josephson junction, this effect is not yet understood but explained by us below.

{\it Dynamical washboard potential model.}---
The microscopic theory of Josephson tunnel junctions was established within the tunneling Hamiltonian formalism~~\cite{Cohen,Josephson1962,Ambegakokar1963} and developed to relate current and voltage across the junction to an arbitrary superconducting phase difference $\phi(t)$~\cite{Werthamer,Larkin}. For a DC current bias $I$, the dynamics of $\phi(t)$ follows the nonlinear integro-differential equation, 
\begin{eqnarray}
I = \frac{\hbar}{2eR_n}\frac{d\phi}{dt} &+& \int_{-\infty}^t dt'\mathcal{K}_n(t-t')\sin\frac{\phi(t)-\phi(t')}{2} \nonumber\\
 &+& \int_{-\infty}^t dt'\mathcal{K}_s(t-t')\sin\frac{\phi(t)+\phi(t')}{2}.  \label{Eq:EoM}
\end{eqnarray}
The first term on the right hand side is the instantaneous Ohmic response of the quasiparticle current. The second term is the retarded response of the quasiparticle current. 
The third term is the retarded response of the supercurrent. The retarded responses stem from the frequency-dependent tunneling currents caused by the particular gap structures of given superconducting electrodes~\cite{Barone,Tafuri}. Memory kernels $\mathcal{K}_{n,s}(t)$ describe retarded responses by coupling the past dynamics of $\phi(t')$ at $t'$ to $\phi(t)$ at the present time $t>t'$. The dynamics of $\phi(t)$ determines the DC voltage drop $V\equiv\lim_{\tau\rightarrow\infty}\frac{1}{\tau}\int_0^\tau dt\,v(t)$ across the junction with $v(t)=\frac{\hbar}{2e}\frac{d\phi}{dt}$.

We recast Eq.~\eqref{Eq:EoM} into a novel form of the dynamical washboard potential model with nonequilibrium modulations of certain parameters,
\begin{equation}
I = \frac{\hbar}{2eR_n}\frac{d\phi}{dt} + J(t)\sin[\phi(t)-\zeta(t)] + S(t). \label{Eq:DynamicWashboard}
\end{equation}
Due to the nonlocality-in-time of Eq.~\eqref{Eq:EoM}, the phase difference in time $\phi(t)-\phi(t')$ causes the nonequilibrium modulations $I_{s,n}^\text{neq}(t)\equiv\int_{-\infty}^t dt'\,\mathcal{K}_{s,n}(t-t')e^{i\frac{\phi(t)-\phi(t')}{2}}$. $J(t)$ and $\zeta(t)$ are, respectively, amplitude and argument of the nonequilibrium modulations of the supercurrent $I_s^\text{neq}(t)$, while $S(t)$ is the imaginary part of the quasiparticle current $I_n^\text{neq}(t)$. The reformulation into Eq.~\eqref{Eq:DynamicWashboard} enables us to develop a qualitative analysis using the mechanical analogue of the so-called phase particle in a tilted washboard potential $\mathcal{U}(\phi,t)=[S(t)-I]\phi-J(t)\cos[\phi-\zeta(t)]$ and a quantitative analysis employing an iterative approach.

\begin{figure}[b]
\centering
\includegraphics[width=0.99\columnwidth]{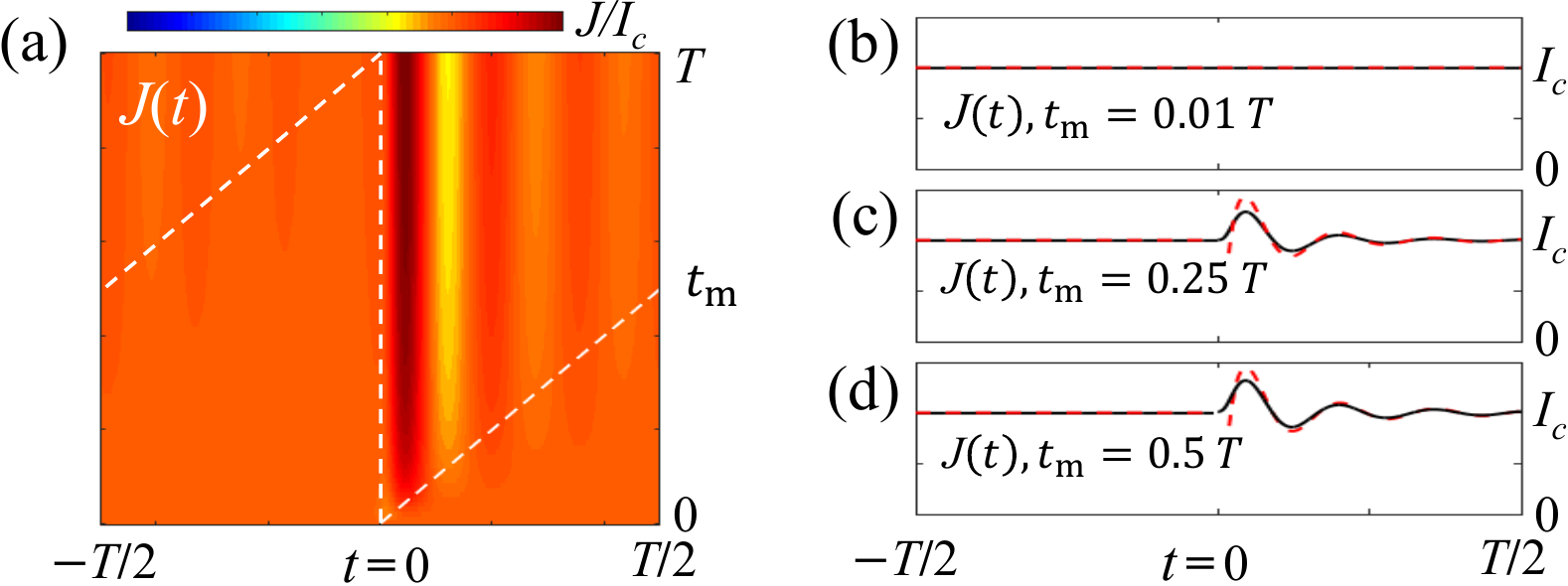}
\caption{Analytical estimation and numerical calculation of the modulating height $J(t)$ of the dynamical washboard potential $\mathcal{U}(\phi,t)$. We depict a numerically calculated color map of $J(t)$ with respect to the memory time $t_\text{m}$ in (a). The time ranges where the voltage pulse modulates $\mathcal{U}(\phi,t)$ are indicated as a guide for the eye (dashed lines). In the right panels, we present plots of $J(t)$ for several $t_\text{m}$ for a better comparison of analytics (dashed lines) to numerics (solid lines). The height $J(t)$ of $\mathcal{U}(\phi,t)$ oscillates around $I_c$. We use $\phi_0(t)$ exhibiting $\delta T/T=0.05$ and the memory kernels from Ref.~\cite{Harris1976}. Note that $\mathcal{U}(\phi,t+T)=\mathcal{U}(\phi,t)$ and $\phi(t+T)=\phi(t)+2\pi$~\cite{Periodicity}.
} \label{Fig1}
\end{figure}

{\it Qualitative analysis.}---The mechanical analogue of Eq.~\eqref{Eq:DynamicWashboard} allows a qualitative analysis of $\phi(t)$ without seeking the exact solution. 
When $I_c>I$, the dynamics becomes stationary with $\phi(t)-\phi(t')=0$ converging to the fixed point $\phi_\times=\arcsin(I/I_c)$. This results in $V=0$ and $I_c=\int_0^\infty\,dt\,\mathcal{K}_{s}(t)$.
When $I_c<I$, $\phi(t)$ evolves in absence of a fixed point. Then, the phase difference in time modulates $\mathcal{U}(\phi,t)$.
To analyze this complex behavior, we add factors $e^{-t/t_\text{m}}$ to the memory kernels $\mathcal{K}_{s,n}(t)$ by hand. These factors introduce a finite memory time $t_\text{m}$. 
We qualitatively analyze the influence of the past dynamics on $\mathcal{U}(\phi,t)$ and $\phi(t)$, starting with an exactly solvable limit $t_\text{m}\rightarrow0$ (no memory). Subsequently, we turn on the nonlocality-in-time by increasing $t_\text{m}$. In the end, this gives rise to a qualitative understanding of the $I$-$V$ characteristics when $I_c<I$.

The limit $t_\text{m}\rightarrow0$ yields the RSJ model where Eq.~\eqref{Eq:DynamicWashboard} becomes local-in-time and exactly solvable: $\phi(t)\rightarrow\phi_0(t)=2\arctan[\frac{I_c}{I}+\frac{\sqrt{I^2-I_c^2}}{I}\tan\frac{eR_nt\sqrt{I^2-I_c^2}}{\hbar}]$ and $\mathcal{U}(\phi,t)\rightarrow\mathcal{U}_0(\phi)=-I\phi-I_c\cos\phi$ with $J_0(t)=I_c$ and $\zeta_0(t)=S_0(t)=0$, yielding the $I$-$V$ characteristics $V=R_n\sqrt{I^2-I_c^2}$. In the low voltage regime $V\ll I_cR_n$, $\phi_0(t)$ shows an abrupt phase winding by $2\pi$ for every period as a step-wise jump, generating sharp voltage pulses with $\max\{v(t)\}\sim I_cR_n$ and time width $\delta T\sim\hbar/(eI_cR_n)$. 

Next, we consider the memory effect with finite $t_\text{m}<T$ and analyze the behavior of the tilted washboard potential $\mathcal{U}(\phi,t)$. We begin this analysis for pedagogical reasons by focusing on a single voltage pulse of $\phi_0(t)$ appearing at $t=0$ with a short time width $\delta T\ll T$. Then, sufficiently long after or before the voltage pulse, $\mathcal{U}(\phi,t)$ hardly varies from $\mathcal{U}_0(\phi)$ owing to the finite memory time $t_\text{m}$ and causality. However, in the vicinity of the voltage pulse $0<t<t_\text{m}$, we argue that the abrupt $2\pi$ phase winding  modulates $\mathcal{U}(\phi,t)$ with $I_{s,n}^\text{neq}(t)\approx\int_{-\infty}^0 dt'\mathcal{K}_{s,n}(t-t')e^{\frac{t'-t}{t_\text{m}}}-\int_{0}^t dt'\mathcal{K}_{n,s}(t-t')e^{\frac{t'-t}{t_\text{m}}}$ using the expression below Eq.~\eqref{Eq:DynamicWashboard}~\cite{footnote}.
The nonequilibrium modulations of $\mathcal{U}(\phi,t)$ are determined by the memory kernels $\mathcal{K}_{n,s}(t)$ via $J(t)\approx I_c -  2\int_{t}^{t+t_\text{m}}d\tilde{t}\,\mathcal{K}_s(\tilde{t})e^{-\tilde{t}/t_\text{m}}$, $\zeta(t) \approx \,\mathcal{K}_s(t)e^{-t/t_\text{m}}\delta T/I_c$, and $S(t)\approx -\,\mathcal{K}_n(t)e^{-t/t_\text{m}}\delta T$. Hence, $\mathcal{U}(\phi,t)$ oscillates around $\mathcal{U}_0(\phi)$. We compare our estimation of $J(t)$ with exact numerical calculations for a tunnel junction coupling two BCS superconductors in Fig.~\ref{Fig1}. For this type of junction, $\mathcal{K}_{n,s}(t)$ sinusoidally oscillate with the time scale $\hbar/(2\Delta)$ due to energy-time uncertainty, and $\mathcal{K}_{n,s}(t)\rightarrow0$ for $t\rightarrow\infty$ due to causality~\cite{Harris1976}.

We can derive the general behavior of $\phi(t)$ from the mechanical analogue of a phase particle in presence of a finite memory time $t_\text{m}<T$. 
According to the modulation of $\mathcal{U}(\phi,t)$ around $\mathcal{U}_0(\phi)$, the phase particle exhibits back-and-forth motion. This can be described by an additional function $\delta\phi(t)$, which oscillates around zero. After several back-and-forth motions, the phase particle $\phi(t)$ eventually advances by $2\pi$ due to the lack of a fixed point for $I_c<I$. We replace $\phi(t)\rightarrow\phi_0(t)+\delta\phi(t)$ including a memory effect. As we increase the nonlocality-in-time further $T<t_\text{m}$,  more voltage pulses in the past contribute to the modulation of $\mathcal{U}(\phi,t)$. 

We develop the following picture of the memory effect on the $I$-$V$ characteristics: $I$ corresponds to the tilting of the dynamical washboard potential and $V$ to the inverse of the time period $T=\pi\hbar/(eV)$ satisfying $\phi(t+T)=\phi(t)+2\pi$.
Let us assume that a static washboard potential $\mathcal{U}_0(\phi)$ should be tilted by $I_0$ in order to generate a voltage drop $V=\pi\hbar/(eT)$. Then, the memory effect demands larger tilting $I$ of the dynamical washboard potential $\mathcal{U}(\phi,t)$ than $I_0$, since the memory effect makes it more difficult for the phase particle to advance by $2\pi$ because of the back-and-forth motion. Hence, the memory effect changes the $I$-$V$ characteristics, accordingly.

{\it Quantitative analysis.}---We now present an analytical approach for the DC current-biased $I$-$V$ characteristics of a generic Josephson tunnel junction. The task is to calculate the DC current-bias $I$ generating the DC voltage drop $V$. We approach this task by an iterative solution of Eq.~\eqref{Eq:DynamicWashboard}. The $N$-th iteration step begins with calculating the dynamical modulations of $\mathcal{U}_N(\phi,t)$ using the previous iterative solution $\phi_{N-1}(t)$ with the period $T=\pi\hbar/(eV)$. The $N$-th iterative solution $\phi_N(t)$ is obtained by solving Eq.~\eqref{Eq:DynamicWashboard} with $\mathcal{U}_N(\phi,t)$. Those iterations are repeated until the solution converges. Intriguingly, our quantitative analysis shows that the first iteration step is sufficient to obtain the $I$-$V$ characteristics to a good accuracy.

The reason for the power of the first iteration step is related to a clever choice of the initial ansatz $\phi_0(t)=2\arctan[\frac{I_cR_n+V\tan(eVt/\hbar)}{\sqrt{(I_cR_n)^2+V^2}}]$, inspired by the functional form of the solution of the RSJ model. Updating $J_0(t)=I_c$, $\zeta_0(t)=S_0(t)=0$ into the dynamical modulations $J_1(t)$, $\zeta_1(t)$, $S_1(t)$ by inserting $\phi_0(t)$ into the expressions defined below Eq.~\eqref{Eq:DynamicWashboard}, we take the memory effect into account. The first iterative solution $\phi_1(t)$ is obtained from
\begin{equation}
I = \frac{\hbar}{2eR_n}\frac{d\phi_1}{dt} + J_1(t)\sin[\phi_1(t)-\zeta_1(t)] + S_1(t). \label{Eq:FirstIteration}
\end{equation}
$\phi_1(t)$ exhibits back-and-forth motion with the correction $\delta\phi_1(t)=\phi_1(t)-\phi_0(t)$. We reduce Eq.~\eqref{Eq:FirstIteration} into the equation of motion of $\delta\phi_1(t)$ in the regime $J_1(t)/I_c-1,\zeta_1(t),S_1(t)/I_c\ll 1$. This implies that $\mathcal{U}_1(\phi,t)$ oscillates weakly around $\mathcal{U}_0(\phi)$. The reduced equation of motion describes the dynamics of $\delta\phi_1(t)$ in a different washboard potential with the height $I_c$ but dynamically tilted by $[I_c-J_1(t)]\sin\phi_0(t)- S_1(t)$. Since the tilting is much smaller than its height, $\delta\phi(t)$ lacks the $2\pi$ phase winding and shows weak oscillations around zero with a small DC component. The iteration steps can be repeated, but the corrections due to additional small oscillations are negligible.

This motivates us to use the first iteration to calculate the DC current-bias $I$ generating the DC voltage drop $V$ and analyze how well it works. We put $\phi(t)=\phi_0(t)+\delta\phi(t)$ with a small oscillating correction $\delta\phi(t)$ into Eq.~\eqref{Eq:DynamicWashboard}, letting $\phi(t)$ and $\phi_0(t)$ share the period $T=\pi\hbar/(eV)$ producing the DC voltage drop $V$ with $\delta\phi(t+T)=\delta\phi(t)$. Collecting terms containing other than $\phi_0(t)$ as $\mathcal{O}[\delta\phi(t)]$, we rewrite Eq.~\eqref{Eq:DynamicWashboard} as
\begin{equation}
I = \frac{\hbar}{2eR_n}\frac{d\phi_0}{dt} + J_1(t)\sin[\phi_0(t)-\zeta_1(t)] + S_1(t) + \mathcal{O}[\delta\phi(t)].  \label{Eq:ModifiedWP_phi0}
\end{equation}
To evaluate the DC current-bias $I$, we take the time-average of Eq.~\eqref{Eq:ModifiedWP_phi0} over the period $T=\pi\hbar/(eV)$. Then, the time integral of $\mathcal{O}[\delta\phi(t)]$ contains an integrand multiplied by $\delta\phi(t)$. Using $\delta\phi(t+T)=\delta\phi(t)$, we find $\int_0^T dt\,f(t)\delta\phi(t) \lesssim \frac{A_m}{2m\pi}\int_0^T dt\,f(t)$, where $m$ and $A_m$ are, respectively, the number of back-and-forth motions within a period and the amplitude of $\delta\phi(t)$. Since $A_m$ is small and $m>1$, we find that the time-average of $\mathcal{O}[\delta\phi(t)]$ becomes negligibly smaller than $I_c$. Thus, we obtain this equation for the $I$-$V$ characteristics
\begin{equation}
I \approx \frac{V}{R_n} + \overline{J_1(t)\sin[\phi_0(t)-\zeta_1(t)]} + \overline{ S_1(t)}, \label{Eq:SelfConsistency}
\end{equation}
where $\overline{f(t)}\equiv\frac{1}{T}\int_0^T dt\,f(t)$. 

Focusing on the low-voltage regime $V<I_cR_n$, by which we can approximate $\phi_0(t)$, we evaluate the time-averages in Eq.~\eqref{Eq:SelfConsistency} further. This yields an analytical formula for the DC current-biased $I$-$V$ characteristics of Josephson tunnel junctions with generic memory kernels,
\begin{eqnarray}
I \approx && \sqrt{I_c^2+\left(\frac{V}{R_n}\right)^2} + \frac{\hbar}{eI_cR_n}\sum_{n=0}^\infty(-1)^n \overline{\mathcal{K}_n\left(t+n\frac{\pi\hbar}{eV}\right)} \nonumber\\
&& - 2\sum_{n=0}^\infty(-1)^n \int_{-\infty}^{-n\frac{\pi\hbar}{eV}}dt'\,\overline{\mathcal{K}_s(t-t')}. \label{Eq:CVC}
\end{eqnarray}
This equation is the first main result of our work. The additional contributions of the second and third terms in Eq.~\eqref{Eq:CVC} stem from the memory effects of the quasiparticle current and supercurrent at the current bias. Since memory kernels contain the retarded dynamics of electric fields from voltage pulses, the time-averaged memory kernels are the central physical quantities determining the $I$-$V$ characteristics at low voltages. The summations signify the influence of each voltage pulse in the past on the dynamical washboard potential. We discuss the opposite limit, i.e., the $I$-$V$ characteristics of Josephson tunnel junctions in the high-voltage regime $I_cR_n< V$, in the Supplemental Material by employing Eq.~\eqref{Eq:SelfConsistency}~\cite{SM}.

{\it Werthamer equation with smeared DOS.}---We exemplify the validity of the analytical formula, Eq.~\eqref{Eq:CVC}, by looking at a concrete example of a Josephson tunnel junction. The junction consists of two BCS superconductors with a small junction cross section. The Werthamer equation describes the Josephson effect by taking into account the memory kernels of the junction in Eq.~\eqref{Eq:EoM}. It shows good agreement with experiments, when smearing of the superconducting gap is considered~\cite{Zorin,Likharev}. The smearing appears in realistic situations due to various mechanisms~\cite{Mourik,Das,Deng,Finck,Zorin1979,Zorin1979,Likharev}. 
Considering the smeared DOS, we derive an analytical formula of the $I$-$V$ characteristics and compare it to numerics~\cite{McDonald, Schlup1978, Zorin}. We show below that different types of the $I$-$V$ characteristics are interpolated within our theoretical framework as the smearing increases.

We take into account the smeared DOS with an energy broadening $\Gamma$ to describe BCS superconductors. This gives rise to memory kernels $\mathcal{K}_s(t) = -\pi\Delta^2e^{-\Gamma t/\hbar}J_0(t\Delta/\hbar)Y_0(t\Delta/\hbar)/(\hbar e R_n)$ and $\mathcal{K}_n(t) = \pi\Delta^2e^{-\Gamma t/\hbar}J_1(t\Delta/\hbar)Y_1(t\Delta/\hbar)/(\hbar e R_n)$ at zero temperature. $J_n$ and $Y_n$ are Bessel functions of first and second kinds, respectively. Symmetric superconducting gaps $\Delta$ are considered for both superconductors for simplicity. Putting the above memory kernels into Eq.~\eqref{Eq:CVC}, we derive the DC current-biased $I$-$V$ characteristics in the low-voltage regime in analytical form, which is the second main result of our work,
\begin{widetext}
\begin{equation}
\frac{I}{I_c} \approx \sqrt{1+\left(\frac{V}{I_cR_n}\right)^2} - \frac{4V}{\pi^2 V_g} + \frac{4\arctan\left(\frac{\sin\frac{\pi V_g}{V}}{e^{\pi\gamma V_g/V}+\cos\frac{\pi V_g}{V}}\right)}{\pi^2(1+\gamma^2)[K(-\gamma^2)]^2}\frac{V^2}{V_g^2} - \frac{4\log\sqrt{2e^{-\frac{\pi\gamma V_g}{V}}\left(\cosh\frac{\pi\gamma V_g}{V}+\cos\frac{\pi V_g}{V}\right)}}{\pi^2(1+\gamma^2)K(-\gamma^2)}\frac{V^2}{V_g^2}, \label{Eq:IVwerthamer}
\end{equation}
\end{widetext}
where $\gamma=\Gamma/(2\Delta)$ and  $V_g=2\Delta/e$. The critical current $I_c=\Delta/(eR_n)K(-\gamma^2)$ decreases as the smearing energy increases. $K(x)$ is the complete elliptic integral of the first kind. We note that the last two terms on the right hand side in Eq.~\eqref{Eq:IVwerthamer} stem from the retarded responses to voltage pulses in the past. The third term corresponds to the step-wise increasing quasiparticle current by photon-assisted tunneling~\cite{McDonald}. The fourth term describes the subharmonic peaks of the supercurrent by self-coupling~\cite{Werthamer,McDonald}. 

\begin{figure}[b]
\centering
\includegraphics[width=0.99\columnwidth]{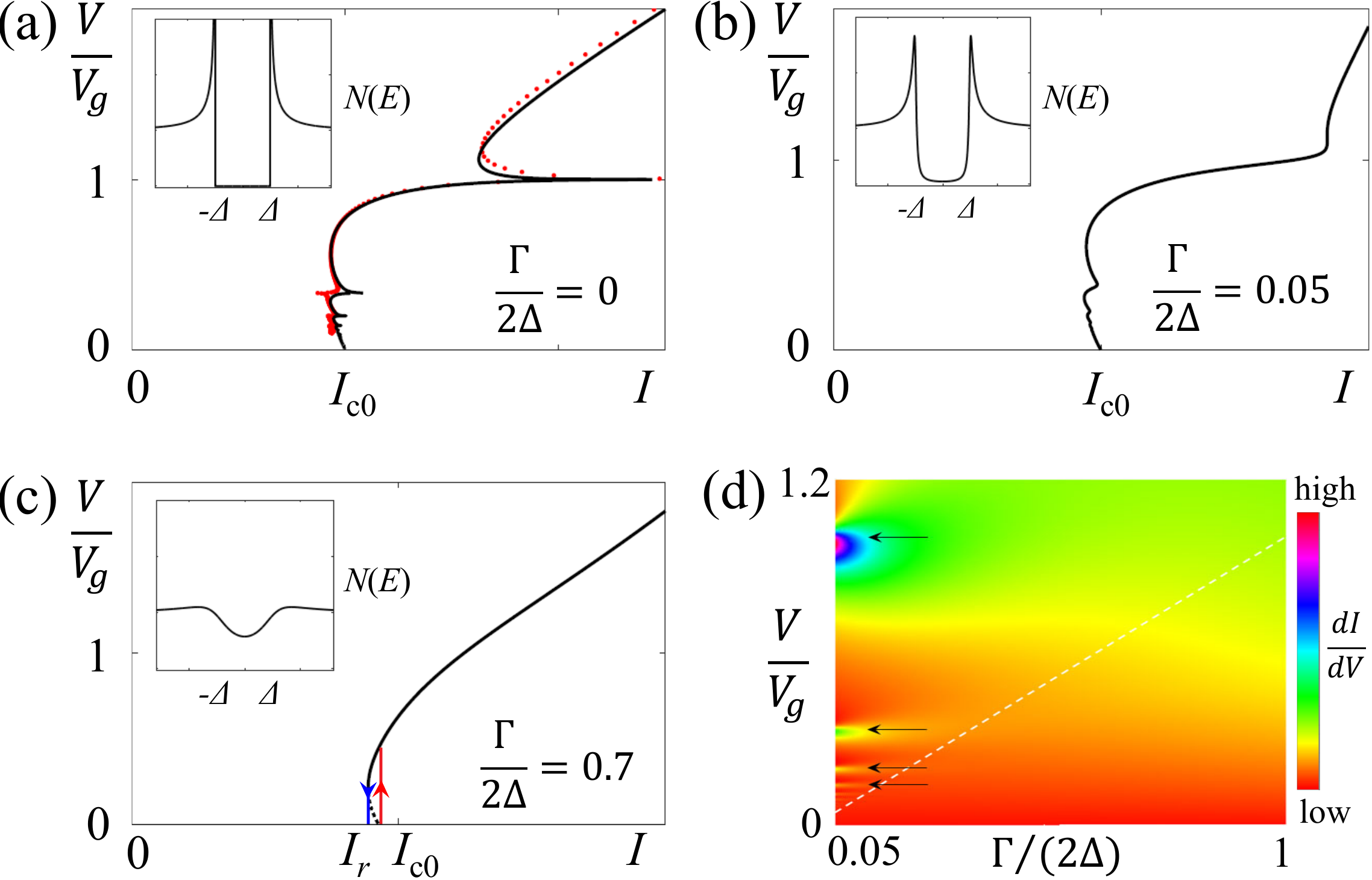}
\caption{DC current-biased $I$-$V$ characteristics based on the Werthamer equation with smeared DOS $N(E)$. In (a), our analytical result (solid line) is provided together with a numerical calculation (red dots) using Eq.~\eqref{Eq:EoM} without smearing. Analytical calculations for $\Gamma\ll\Delta$ and $\Gamma\sim\Delta$ are shown in (b) and (c), respectively. The voltage staircases become visible as peaks in the colormap of $dI/dV$ in (d). $V/V_g=1,1/3,1/5,1/7$ are indicated with arrows from top to bottom. $V=\Gamma/e$ is drawn as a guide to the eye (dashed line) and $I_{c0}=\pi\Delta/(2eR_n)$. Insets: Quasiparticle DOS $N(E)$ corresponding to the chosen $\Gamma$ in each panel~\cite{SM}. 
} \label{Fig2}
\end{figure}

We begin with the case without smearing, which has been studied extensively with numerical approaches~\cite{McDonald, Schlup1978}. When $\Gamma=0$, Eq.~\eqref{Eq:IVwerthamer} reproduces the logarithmic divergence in the subharmonics at voltages $V_p=V_g/(2p-1)$ ($p=1,2,\dots$) and $I_cR_n=\pi\Delta/(2e)$~\cite{Werthamer,McDonald}. We compare the analytical formula with numerics in Fig.~\ref{Fig2}(a). Along with the notable agreement in both calculations, our analytical approach clarifies that the logarithmic divergence stems from the infinitely large number of voltage pulses due to the algebraically decaying memory kernels $\mathcal{K}_{n,s}(t)\sim1/t$ for $\Gamma=0$. This behavior suggests that the logarithmic divergence is washed out and limited by a finite measurement time in experiments.

For small smearing $\Gamma\ll\Delta$, Eq.~\eqref{Eq:IVwerthamer} implies the emergence of voltage staircases in the absence of AC current bias, reported in recent experiments with high-quality Josephson junctions~\cite{Gul,Charpentier,Mayer,Ridderbos,Perla}. Notably, it is theoretically suggested that the dynamics of a qubit in junctions can replace the role of the AC current bias employing the quantum RSJ model~\cite{Choi2020,Oriekhov2021}. However, our theoretical analysis shows that voltage staircases can even appear in absence of both AC current bias and qubit dynamics in the junction. We discover that the feedback effect modulates the washboard potential dynamically with $[S(t)-I]\phi-J(t)\cos[\phi-\zeta(t)]$. This feedback effect becomes most significant when voltage pulses add constructively $V_p=V_g/(2p-1)$ ($p=1,2,\cdots$). The resulting dynamical modulation causes a similar behavior found in the quantum RSJ model with qubit dynamics~\cite{Feng2018,Choi2020,Oriekhov2021}. Then, the phase particle is trapped against increased tilting by a current bias, which results in rather flat voltage drops at $V_p$ [see Figs.~\ref{Fig2}(b) and \ref{Fig2}(d)].

Moreover, the comparison of the time scales of memory kernels provides further understanding, for instance, related to its oscillation period $T_\Delta=\hbar/(2\Delta)$ and decaying time scale $T_\Gamma=\hbar/\Gamma$. The exponential decay with $T_\Gamma=\hbar/\Gamma$ explains the absence of the logarithmic divergence occurring for $\Gamma=0$. The decaying time $T_\Gamma$ competes with the period of voltage pulses $T=\pi\hbar/(eV)$, as less contributions from the past add together at present with increasing period $T$ between voltage pulses for a given $T_\Gamma$.  Consequently, the voltage staircases appear approximately in the range $\Gamma/e<V<2\Delta/e$ [see Fig.~\ref{Fig2}(d)].

Finally, we consider the regime of large smearing $2\Delta\lesssim\Gamma$, at which the memory kernel shows over-damped oscillations. Then, the contributions from retarded responses, the last two terms in Eq.~\eqref{Eq:IVwerthamer}, become vanishingly small, and the $I$-$V$ characteristics at zero temperature qualitatively coincides with the one from the RSJ model  [Fig.~\ref{Fig2}(c)]. However, we find that the nonequilibrium Josephson effect causes additional hystereses in the $I$-$V$ characteristics. Eq.~\eqref{Eq:IVwerthamer} yields the recapturing current $I_r=I_c\sqrt{1-4[K(-\gamma^2)]^2/\pi^4}$ where the voltage drop vanishes at backward current bias. The hysteresis emerges as our theory can capture the quasiparticle excitations and stored Josephson energy.

{\it Summary}---We provide a generic theory of the DC current-biased $I$-$V$ characteristics applicable to various types of Josephson tunnel junctions.  In our theory, the information of the junction is contained in memory kernels. We derive an analytical formula which can be used as a powerful fitting function to experiments under appropriate conditions. In sharp contrast to the voltage-biased case, where multiple Andreev reflections of quasiparticles are responsible for subharmonics, we show that the supercurrent can play a major role in the DC current-biased $I$-$V$ characteristics.  Our analytical formula can be employed to quantitatively evaluate the shape of the superconducting gap. It also explains the hystereses experimentally observed in low-dimensional Josephson junctions where the geometric capacitance is negligibly small.

\begin{acknowledgments}
This work was supported by the W\"urzburg-Dresden Cluster of Excellence on Complexity and Topology in Quantum Matter (EXC2147, project-id 390858490) and by the DFG (SPP1666 and SFB1170 ``ToCoTronics'').
\end{acknowledgments}

\pagebreak
\widetext
\begin{center}
\textbf{\large Supplemental Material: Microscopic theory of the current-voltage characteristics of Josephson tunnel junctions}
\end{center}

\section{DC current-biased $I$-$V$ characteristics in the high-voltage regime $I_cR_n<V$}\label{Sec:IVhigh}

We provide the $I$-$V$ characteristics of generic Josephson tunnel junctions in the high-voltage regime $I_cR_n< V$. We find that the ansatz in the main text can be written in the high-voltage regime as
\begin{equation}
\phi_0(t)= 2\arctan\left[\frac{I_cR_n+V\tan(eVt/\hbar)]}{\sqrt{(I_cR_n)^2+V^2}}\right] = \frac{2eV}{\hbar}t + 2\frac{I_cR_n}{V}\cos^2\left(\frac{eVt}{\hbar}\right)+\mathcal{O}\left[\left(\frac{I_cR_n}{V}\right)^2\right].
\end{equation}
By using $\phi_0(t)\approx 2eVt/\hbar + 2(I_cR_n/V)\cos^2(eVt/\hbar)$ with the self-consistent equation Eq.~(4) in the main text, we obtain the $I$-$V$ characteristics in the high-voltage regime $I_cR_n< V$ for generic Josephson tunnel junctions,
\begin{equation}
I\approx\frac{V}{R_n} + \int_0^\infty d\tilde{t}\mathcal{K}_n(\tilde{t})\sin\frac{eV\tilde{t}}{\hbar} + \frac{I_cR_n}{2V}\int_0^\infty d\tilde{t}\mathcal{K}_s(\tilde{t})\cos\frac{eV\tilde{t}}{\hbar}. \label{Eq:CVChigh}
\end{equation}
The quasiparticle current, the first two terms, coincides with the $I$-$V$ characteristics of the DC voltage-biased case $V$ due to the similarity of the dynamics of $\phi_0(t)$ to that of the DC voltage-biased case $\phi_0(t)=2eVt/\hbar$. However, the supercurrent shows the difference in the $I$-$V$ characteristics depending on the type of bias. Owing to the additional oscillating dynamics $2(I_cR_n/V)\cos^2(eVt/\hbar)$ in case of the DC current bias, the supercurrent in the third term carries a non-vanishing DC component, which affects the $I$-$V$ characteristics. In contrast, the DC voltage-biased $I$-$V$ characteristics only include sinusoidal AC supercurrents.

Employing the memory kernels in the Werthamer equation without smearing energy $\Gamma$, we obtain the $I$-$V$ characteristics in the high-voltage regime,
\begin{equation}
I(V)\approx \frac{V+V_g}{R_n}E\left[\left(\frac{V-V_g}{V+V_g}\right)^2\right] - \frac{V_g(2V+V_g)}{R_n(V+V_g)}K\left[\left(\frac{V-V_g}{V+V_g}\right)^2\right] + I_c\frac{V_g^2}{4V^2}K\left(\frac{V_g^2}{V^2}\right), \label{Eq:WertahmerHigh}
\end{equation}
where $I_cR_n=\pi\Delta/(2e)$ for $\Gamma=0$. $K(x)$ and $E(x)$ are the complete elliptic integrals of the first and second kind, respectively. Note that the DC supercurrent in the last term is responsible for the logarithmic divergence at $V=V_g$, which is absent in the DC voltage-biased $I$-$V$ characteristics. We use Eq.~\eqref{Eq:WertahmerHigh} for Fig.~2(a) in the main text.

\section{Smeared density of states of BCS superconductors}\label{Sec:DOS}
We derive the density of states with the smearing parameter $\Gamma$. The density of states is calculated by
\begin{equation}
N(E) = -\frac{1}{\pi\mathcal{V}}\sum_{\mathbf{k}}\text{Im}\{G(\mathbf{k},E+i\Gamma)\} = -\frac{1}{\pi\mathcal{V}}\sum_{\mathbf{k}}\text{Im}\left\{\frac{E+i\Gamma+\xi_\mathbf{k}}{(E+i\Gamma)^2-E_\mathbf{k}^2}\right\}.
\end{equation}
We use the normal Green function of BCS superconductors $G(\mathbf{k},E+i\Gamma)$ with the smearing energy $\Gamma$ added as an imaginary part to the energy. $E_\mathbf{k} = \sqrt{\xi_\mathbf{k}+\Delta^2}$ is the single particle excitation spectrum of the Bogoliubov quasiparticles. $\mathcal{V}$ is the volume of the system. By changing the summation over discrete wave numbers $\mathbf{k}$ to an integration over the continuous energy $\xi$, the smeared density of states is calculated as
\begin{equation}
N(E) = N(0)\text{Re}\left\{\frac{|E|-i\Gamma}{\sqrt{(|E|+i\Gamma)^2-\Delta^2}}\right\}.
\end{equation}
We note that the BCS DOS $N(E)=N(0)\Theta(|E|-\Delta)|E|/\sqrt{E^2-\Delta^2}$ is obtained in the limit $\Gamma\rightarrow0$. $\Theta(x)$ is the Heaviside step function.
The smeared DOS $N(E)$ is in good agreement with experimental measurements of the DOS using tunneling spectroscopy~\cite{Dynes1978,Dynes1984}.


\begin{thebibliography}{70}
\bibitem{footnote} The $2\pi$ phase winding has flipped the sign of the second term and resulted in the additional modulation $-2\int_{0}^t dt'\mathcal{K}_{n,s}(t-t')e^{\frac{t'-t}{t_\text{m}}}$, compared to the case without the phase winding.
\end{thebibliography}
\end{document}